\documentclass[aps,twocolumn,showpacs]{revtex4}
\usepackage{graphicx}

\begin{document}
%\draft

%<<<<<<<<<<<<< TITLE >>>>>>>>>>>>>>>%
\title{Energy density bounds for black strings }

%<<<<<<<<<<<<< AUTHOR >>>>>>>>>>>>>>>%
\author{Shinya Tomizawa\footnote{tomizawa@sci.osaka-cu.ac.jp}
 }

%<<<<<<<<<<<<< ADDRESS >>>>>>>>>>>>>>>%
\affiliation{Department of Physics, Osaka City University, 3-3-138, Sugimoto, Sumiyoshiku, Osaka City, Osaka, 113-0033, Japan}

%<<<<<<<<<<<<< DATE >>>>>>>>>>>>>>>%
\date{\today}

%======================================%
%<<<<<<<<<<<<< ABSTRACT >>>>>>>>>>>>>>>%
%======================================%
\begin{abstract}
The conserved charge called Y-ADM mass density associated with asymptotically translational Killing-Yano tensor gives us an appropriate physical meaning about the energy density of $p$ brane spacetimes or black strings. We investigated the positivity of energy density in black string spacetimes, using the spinorial technique introduced by Witten. Recently, the positivity of Y-ADM mass density in $p$ brane spacetimes was discussed. In this paper, we will extend this discussion to the transversely asymptotically flat black string spacetimes containing an apparent horizon. We will give the sufficient conditions for the Y-ADM mass density to become positive in such spacetimes.

\end{abstract}


\maketitle
%\vskip1cm

%======================================%
%<<<<<<<<<<<< SECTION I  >>>>>>>>>>>>>>%
%======================================%
%\baselineskip25pt
\section{Introduction}
\label{sec:introduction}
One of most important and interesting properties of black strings is the classical instability  which was discovered by Gregory and Laflamme \cite{GL}. However, at this stage, we have not obtain the clear answers for what is the final state of the instability. However, the positivity of the energy of the spacetimes, that is, whether there is the lower bound of the energy of the spacetimes, or not,  might give us the information about the stability of them or the final states. For an example, the Schwarzschild black hole with negative mass is unstable under the metric perturbation, and if the positivity of the energy holds, then the spacetimes with negative mass would have naked singularity somewhere.  A sufficient condition for the stability of spacetimes is that its energy saturates a lower bound on the energy of all field configurations satisfying the same boundary conditions. It is known that some charged extreme black hole solutions saturate a lower bound such a bound, which is the Bogomol' nyi-type bound \cite{GH}.

In a class of four dimensional asymptotically flat spacetimes, the trial to show the positivity of the ADM energy had been made of by a lot of authors. However, they proved it in the only special cases.
Finally, Shoen and Yau \cite{SY} succeeded in giving a complete proof on the positivity of the ADM energy in asymptotically flat spacetimes, which can be applied whether there is a black hole or not. The more elegant and simpler proof was given by Witten and Nester \cite{W,N}, who used the spinorial method, when there is no horizon. The introduced spinor is the solution of Dirac-Witten equation and is asymptotically constant at spatial infinity.  Gibbons et. al. \cite{G} extended Witten's arguments to black hole spacetimes containing an apparent horizon without assuming anything about the interior region of a black hole.  They integrated the Dirac-Witten equation over a sapcelike hypersurface with the boundaries the spatial inifinity and the horizon and applied Stokes's theorem there.  The term over the apparent horizon as an inner boundary arise in addition to the surface term over the spatial infinity, which coincides with the ADM 4-momentum. Since this spinor boundary term over the apparent horizon becomes proportional to the expansion of null geodesics on the apparent horizon, it vanishes. As a result, the discussion of Gibbons et. al. becomes equal to that of the original Witten's and the ADM mass also becomes positive in the presence of a black hole.

The ADM mass in a $p$-brane spacetime or a black string will diverge because the spatial $p$ directions or the direction of the black sting are infinite in extent and it is given by an integral over the spatial infinity (with its topology $R^p\times S^{D-(p+2)}$) which encloses the $p$-brane. Therefore, in such a spacetime the ADM mass is not well-defined conserved charge in general except for the cases where they are compactified. The proof of positivity of such compactified cases have been already given \cite{SIT}.  Through this paper, we consider the conserved charges from asymptotic Killing-Yano tensors, which was introduced by Kastor and Traschen \cite{KT}. This charge is called Y-ADM charge and gives us good physical interpretation of the mass density rather than that of a total mass of the spacetime. In the previous work \cite{YADM}, we investigated the positivity of Y-ADM mass density in $1$ brane spacetimes, which are assume to be transverse asymptotically flat to the $1$ brane. As results, we could establish the positivity in the two special cases ; (a) conformally flat cases or algebraically special spacetimes and (b) cases where there exists translational Killing vector field along the string.

In this paper, we will investigate the positivity of the Y-ADM mass density in transversely asymptotically flat black string spacetimes containing an apparent horizon, using Witten-Nester's spinorial technique. Due to the presence of an horizon, the spinor boundary term over the cross section of the horizon and the transverse space arises, as is seen in the proof of Gibbons et. al. \cite{G}. If this boundary term vanishes, we can establish the positivity of the Y-ADM mass density in the above cases (a) and (b) from the results in the previous studies \cite{YADM}. We will show the sufficient conditions for the horizon surface term to vanish, which is the cases where (A) the apparent horizon becomes null. Therefore, if at least, the black string spacetimes satisfy the conditions either (a), or (b) and (A), then the positivity of Y-ADM mass density is assured. 

This paper is organized as follows. In section \ref{sec:surface}, we will show that the spinorial boundary term over the cross section of a black string and the transverse space becomes proportional to the expansion of the cross section rather than the horizon, unlike the proof of Gibbons et. al. \cite{G}. In section \ref{sec:vanish}, we give the sufficient conditions for this surface term to vanish on the apparent horizon. In section \ref{sec:sum}, we will summarize the results.

\section{Positivity bounds of Y-ADM mass density}\label{sec:surface}
We assume that the black string spacetime $(M,g_{ab})$ containing an apparent horizon is $D (\ge 5)$ dimensions and  transverse asymptotically flat, which means asymptotically flat in the direction transverse to the black string but always not in the direction parallel to it. We also assume that the spacetime is foliated by time slices $V_{\rm t}$ with normal timelike vector field $t^a$,  and that each timeslice $V_{\rm t}$ is foliated by the submanifolds $V_{\rm tx}$  binormal to vectors $t^a$ and  $x^a$, where $x^a$ is the vector field pallarel to the black string, satisfying the normalization $x^ax_a=1$ and $t^ax_a=0$. Let us assume the codimension two surface $V_{\rm tx}$ which is the transverse space to the black sting) is foliated by the codimension three surface $V_{\rm txr}$ normal to the transverse radial directional vector field $r^a$ satisfying the $r^ar_a=1$ and $r^at_a=r^ax_a=0$. At transverse spatial infinity we take
$t^a=(\partial/\partial x^0)^a,\ x^a=(\partial/\partial x^1)^a$. In this spacetime, there exists the asymptotic Killing-Yano tensor $f^{(01)}=dx^0\wedge dx^1$, by which we can define the Y-ADM mass density as the surface integral over the transverse spatial infinity $S_{\infty}$ \cite{KT,YADM}, as seen in FIG.1. 

The full spacetime metric can be written in the form

\begin{eqnarray}
g_{ab}&=&-t_at_b+h_{ab}\nonumber\\
      &=&-t_at_b+x_ax_b+q_{ab}\nonumber\\
      &=&-t_at_b+x_ax_b+r_ar_b+r_{ab},
\end{eqnarray}
where $h_{ab}$, $q_{ab}$ and $r_{ab}$ are the induced metrics on the submanifolds $V_{\rm t}$,  $V_{\rm tx}$ and $V_{\rm txr}$, respectively. In this paper, we use the quantities $\gamma^{\hat 0}=-\gamma^at_a,\ \gamma^{\hat 1}=\gamma^ax_a,\ \gamma^{\hat 2}=\gamma^ar_a$.

\begin{figure}[htbp]\label{fig:1}
\begin{center}
\includegraphics[width=0.6\linewidth]{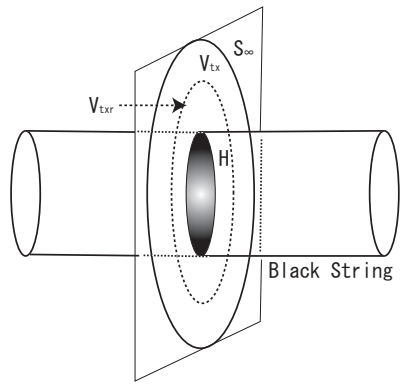}
\end{center}
\caption{$V_{\rm tx}$ is the transverse space which has the boundary $\partial V_{\rm tx}={\cal H}\cup S_{\infty}$, where ${\cal H}$ and $S_{\infty}$ denote the cross section of the black string and transverse space $V_{\rm tx}$ and the transverse spatial infinity, respectively. $V_{\rm tx}$ is foliated by codimension three surfaces $V_{txr}$ which coincide with $S_{\infty}$ on the inner boundary and ${\cal H}$ on the outer boundary.  The transverse space $V_{\rm tx}$ is assumed to be asymptotically flat at the transverse spatial infinity $S_{\infty}$.}
\end{figure}

To prove the positivity of the ADM energy, Witten \cite{W} used the spinorial method, which consists of the two steps. As the first step, he showed that the surface integral of so-called Nester 2 form $E^{ab}=\psi^\dagger\gamma^{\hat 0}\gamma^{abc}\nabla_c\psi$ over the spatial infinity (codimension two surface) coincides with the ADM mass (exactly speaking, ADM $4$-momentum). The following step is to transform this surface integral into the volume integral, using Stokes' theorem,  and to relate it to Eintein tensor equal to the energy momentum tensor satisfying the dominant energy condition.     
To discuss the positivity of Y-ADM mass density, let us introduce the Nester $3$-form \cite{YADM} defined as
\begin{eqnarray}
B^{abc}=\psi^\dagger \gamma^{\hat 0\hat 1}\gamma^{abcd}\nabla_d\psi
\end{eqnarray}
in analogy with the Nester $2$ form in the proof of the positive energy theorem \cite{W}, where $\psi$ is asymptotically constant spinor in the transverse direction and satisfies the Dirac-Witten equation $q^{a}_b\gamma^b\nabla_a\psi=0$.  However, in the reference \cite{YADM}, we dealt with the transversely asymptotically flat $1$ brane spacetimes with no horizon since the in general, in the case with a horizon there exists singularity in the interior region and so the discussion in the reference \cite{YADM} cannot be applied to the spacetimes with horizons, such as, black string. In the latter case, we need add  the surface integral over the cross section $\cal H$ of horizon and transverse space as the inner boundary to the surface integral over the transverse spatial infinity. To explain this, let us begin with the volume integral over the transverse space $V_{\rm tx}$ (See the equation (19) in the reference \cite{YADM}),

\begin{eqnarray}
 &{ }&\frac{1}{8\pi}\int_{V_{tx}}dS_{bc}(\nabla_aB^{abc}+\nabla_aB^{*abc})
                                                   =\frac{1}{8\pi}\int_{V_{\rm tx}}[(G_{ab}t^at^b\nonumber\\&-&R_{abcd}x^ax^cq^{bd} )\psi^\dagger\psi+2(\nabla_a\psi^\dagger)q^{ab}(\nabla_b\psi) ].\label{eq:sur}
\end{eqnarray}

If we choose the cross section of the horizon and the transverse space as an inner boundary in analogy with the reference \cite{G}, then as the result of applying Stokes' theorem to the left hand side in the above eqaution, we obtain the following eqaution,

\begin{widetext}
\begin{eqnarray}
\frac{1}{8\pi}\int_{S_{\infty}}dS_{abc}(B^{abc}+B^{*abc})&-&\frac{1}{8\pi}\int_{\cal H}dS_{abc}(B^{abc}+B^{*abc})\nonumber\\&=&\frac{1}{8\pi}\int_{V_{\rm tx}}[(G_{ab}t^at^b-R_{abcd}x^ax^cq^{bd} )\psi^\dagger\psi+2(\nabla_a\psi^\dagger)q^{ab}(\nabla_b\psi) ],
\end{eqnarray}
\end{widetext}
where the first term in the left hand side coincides with the Y-ADM mass density $\cal M$, following the discussion in the reference \cite{YADM}. On the other hand, the second term in the left hand side appears due to the presence of the horizon $\cal H$, which implies the cross section of the horizon and the transverse space. The right hand side becomes positive in at least, two special cases, (1) conformally flat cases or algebraically special spacetimes and (2) cases where there exists translational Killing vector field along the string, as shown in \cite{YADM}.

The purpose of this paper is to estimate the surface term on the horizon $\cal H$ and to investigate when this surface term vanishes. This term is

\begin{eqnarray}
\frac{1}{8\pi}\int_{\cal H}dS_{abc}(B^{abc}+B^{*abc})=\frac{1}{8\pi}\int_{\cal H}dS(B^{\hat 0\hat 1\hat 2}+B^{\hat 0\hat 1\hat 2}),
\end{eqnarray}
where $B^{\hat 0\hat 1\hat 2}=\psi^\dagger \gamma^{\hat 0}\gamma^{\hat 1}\gamma^{\hat 0\hat 1 \hat 2 }\gamma^{\hat A}\nabla_{\hat A}\psi
                      =\psi^\dagger \gamma^{\hat 2}\gamma^{\hat A}\nabla_{\hat A}\psi$ and $dS$ is the volume element on the cross section of the horizon and the transverse space.

We have the following relation between the projection into $V_t$, $V_{tx}$ and $V_{txr}$ of the $D$ dimensional spinor covariant derivatives $\nabla_a$ and the intrinsic $(D-1$) dimensional, $(D-2)$ dimensional, and $(D-3)$ dimensional spinor covariant derivatives ${}^{(D-1)}\nabla_{\hat A},\ {}^{(D-2)}\nabla_{\hat A}$, and ${}^{(D-3)}\nabla_{\hat A}$, 
\begin{eqnarray}
\nabla_{\hat A}\psi&=&{}^{(D-1)}\nabla_{\hat A}\psi+\frac{1}{2}K_{A\hat i}\gamma^{\hat i}\gamma^{\hat 0}\psi\nonumber\\
                   &=&{}^{(D-2)}\nabla_{\hat A}\psi+\frac{1}{2}J_{\hat A\hat n}\gamma^{\hat n}\gamma^{\hat 1}\psi+\frac{1}{2}K_{A\hat i}\gamma^{\hat i}\gamma^{\hat 0}\psi\nonumber\\
                   &=&{}^{(D-3)}\nabla_{\hat A}\psi+\frac{1}{2}k_{\hat A\hat B}\gamma^{\hat B}\gamma^{\hat 2}\psi+\frac{1}{2}J_{\hat A\hat n}\gamma^{\hat n}\gamma^{\hat 1}\psi\nonumber\\
                   & &+\frac{1}{2}K_{A\hat i}\gamma^{\hat i}\gamma^{\hat 0}\psi,
\end{eqnarray}
respectively, where $K_{ab}$, $J_{ab}$ and $k_{ab}$ are the extrinsic curvatures of the submanifolds $V_{\rm t}$, $V_{\rm tx}$ and $V_{\rm txr}$, respectively. The indexes $\hat i, \hat n$ and $\hat A$ run $\hat i=\hat 1, \cdots, \hat D-1$, $\hat n=\hat 2,\cdots,D-1$ and $\hat A=\hat 3,\dots,\hat,D-1$. Therefore,
the real part of $B^{abc}t_{[a}x_br_{c]}=B^{\hat 0\hat 1 \hat 2}$ becomes
\begin{widetext}
\begin{eqnarray}
B^{\hat 0\hat 1\hat 2}+B^{*\hat 0\hat 1\hat 2}=\psi^\dagger\gamma^{\hat 2}\gamma^{\hat A}{}^{(D-3)}\nabla_{\hat A}\psi+K_{\hat A \hat 2}\psi^\dagger\gamma^{\hat 0}\gamma^{\hat A}\psi+(k^{\hat A}_{\hat A}-K^{\hat 1}_{\hat 1}-K^{\hat 2}_{\hat 2})\psi^\dagger\psi +K^{\hat i}_{\hat i}\psi^\dagger\gamma^{\hat 2}\gamma^{\hat 0}\psi+{\rm c.c}\label{eq:bb}
\end{eqnarray}
\end{widetext}

Let us impose the boundary condition with respect to the spinor on the cross section $\cal H$ of the horizon as follows \cite{G},

\begin{eqnarray}
\gamma^{\hat 2}\gamma^{\hat 0}\psi=\psi,
\end{eqnarray}
where we used the fact that the eigenvalues of the Hermitian matrix $\gamma^{\hat 2}\gamma^{\hat 0}$ are $\pm 1$ and therefore, in the above, we restrict the freedom of the spinor to the half. Under this condition, we find that the first term in the left hand side in the equation (\ref{eq:bb}) vanishes, since 

\begin{eqnarray}
\psi^\dagger\gamma^{\hat 2}\gamma^{\hat A}{}^{(D-3)}\nabla_{\hat A}\psi&=&(\gamma^{\hat 2}\gamma^{\hat 0}\psi)^\dagger\gamma^{\hat 2}\gamma^{\hat A}{}^{(D-3)}\nabla_{\hat A}\psi\nonumber\\
                                                                   &=&-\psi^\dagger\gamma^{\hat 0}\gamma^{\hat 2}\gamma^{\hat 2}\gamma^{\hat A}{}^{(D-3)}\nabla_{\hat A}\psi\nonumber\\
                                                                   &=&-\psi^\dagger\gamma^{\hat 2}\gamma^{\hat A}{}^{(D-3)}\nabla_{\hat A}(\gamma^{\hat 2}\gamma^{\hat 0}\psi)\nonumber\\
                                                                   &=&-\psi^\dagger\gamma^{\hat 2}\gamma^{\hat A}{}^{(D-3)}\nabla_{\hat A}\psi\nonumber\\
                                                                   &=&0.
\end{eqnarray}
Similarly, the second term in the left hand side in the equation (\ref{eq:bb}) also vanishes, because $\gamma^{\hat 0}\gamma^{\hat A}$ anticommute with $\gamma^{\hat 2}\gamma^{\hat 0}$, that is,
\begin{eqnarray}
\psi^\dagger\gamma^{\hat 0}\gamma^{\hat A}\psi=-\psi^\dagger\gamma^{\hat 0}\gamma^{\hat A}\psi=0.
\end{eqnarray} 

Then we obtain the simple expression with respect to the suface integral over the horizon $\cal H$ as follows,
\begin{eqnarray}
& &\frac{1}{8\pi}\int_{\cal H}dS_{abc}(B^{abc}+B^{*abc})\nonumber\\
&=&\frac{1}{8\pi}\int_{\cal H}[\psi^\dagger(k^{\hat A}_{\hat A}-K^{\hat 1}_{\hat 1}-K^{\hat 2}_{\hat 2}+K^{\hat i}_{\hat i})\psi  ],
\end{eqnarray}
which we find that this integrand coincides with the expansion $\theta_+$ of the cross section and black string and the transverse space, since we can observe that the above term coincides with $\theta_+$ defined as  

\begin{eqnarray}
\theta_+&:=&r^{ab}\nabla_al_b\nonumber\\
        &=&\frac{1}{\sqrt{2}}(K^{\hat i}_{\hat i}-K^{\hat 1}_{\hat 1}-K^{\hat 2}_{\hat 2}+k^{\hat A}_{\hat A}),
\end{eqnarray}
where $l^a=\frac{1}{\sqrt{2}}(t^a+r^a)$ is outgoing null vector. Therefore, we observe that when this expansion of the cross section $\cal H$ of the horizon and the transverse space $V_{\rm tx}$, the surface term on the cross section of the horizon vanishes. We should note that unlike the proof in the references \cite{G,SIT},  in general, this term does not vanishes on the apparent horizon, since $\theta_+$ is not expansion of horizon but that of the cross section of the horizon and the transverse space.

\section{Suface term over a horizon}\label{sec:vanish}
As mentioned in the previous section, the surface integral over the cross section of the horizon and the transverse space generally does not vanish on an apparent horizon unlike the proof of Gibbons et. al \cite{G}. As seen in the below, a sufficient for this surface term to vanish is the cases where the apparent horizon becomes null surface. In this section, we will show that the above surface term vanishes on an apparent horizon in the above conditions.

\subsection{Cases where an apparent horizon is null}
Following Hayward's discussion \cite{Hay}, 
an apparent horizon is null if and only if null energy condition hold, the shear and the normal energy density vanishes \cite{Hay} (exactly speaking, this is a necessary and  sufficient condition for a so-called {\it trapping horizon} defined by the closure of a $D-1$ dimensional surface foliated by marginal surfaces on which  $\Theta_+=0, \Theta_-\not=0$ and ${\cal L}_-\Theta_+\not=0$, where ${\cal L}_-$ denotes the Lie derivative along the ingoing null direction and $\Theta_-$ is the expansion of the ingoing null geodesics, becomes null surface. In this paper, the apparent horizon means this trapping horizon. ).
We can show that if an apparent horizon is null surface, the above surface term vanishes. The expansion of out going null geodesics on the apparent horizon is defied as

\begin{eqnarray}
\Theta_+=s^{ab}\nabla_al_b=x^ax^b\nabla_al_b+\theta_+=0\label{eq:ex}\label{eq:expansion}\label{eq:pp}
\end{eqnarray}
On the other hand, the shear of the horizon can be expressed in the form

\begin{eqnarray}
\sigma_{ab}&=&s_{(a}^cs_{b)}^d\nabla_al_b-\frac{1}{D-2}\Theta_+s_{ab}\nonumber\\
           &=&(x^cx^d\nabla_al_b)x_ax_b+(x^c\nabla_cl_d)x_{(a}q_{b)}^d\nonumber\\
           & & +q_{(a}^cq_{b)}^d\nabla_cl_d\nonumber\\
           & &=0,
\end{eqnarray} 
whrere $s_{ab}= g_{ab}+t_at_b-r_ar_b$ and  the equation (\ref{eq:expansion}) was used. Therefore, since $\sigma_{xx}:=\sigma_{ab}x^ax^b=x^cx^d\nabla_cl_d=0$, we can show that on the apparent horizon $\theta_+=0$ in combination with the equation (\ref{eq:ex}).

We will mention when the physical situations where an apparent horizon becomes null would realize. In stationary black string spacetimes, the apparent horizon would coincide with the Killing horizon of the black string. However, even in non-stationary cases, such situation would occur, if there exists the apparent horizon and there is no radiation across and along the horizon. Here we should note that in non-stationary cases, since the apparent horizon is generally inside the horizon, we have to replace $\cal H$ with the cross section $\cal A$ of this apparent horizon $A$ and the transverse as the inter boundary in the equation (\ref{eq:sur}) in order that the inner boundary vanishes.

%\subsection{Translationally symmetric cases}
%In the cases where the there exists a translational Killing vector $x^a$ along %a black string {\it i. e.}, a black string is uniform,  the expansion  of the h%orizon coincides with the expansion of the cross section, that is, $\Theta=\the%ta$, since 

%\begin{eqnarray}
%x^ax^b\nabla_al_b=-x^al^b\nabla_ax_b=x^al^b\nabla_bx_a=0,
%\end{eqnarray}
%where Killing equations (the Killing vector is propotinal to $x^a$) $\nabla_ax_%b+\nabla_bx_a=0$ was used
 %Therefore, in this cases, the surface term over the cross section of the horiz%on and the transvers space vanishes. On the other hand, the positivity of Y-ADM %mass density is shown under the appropriate energy condition \cite{YADM}. As t%hese results, the YADM mass density becomes positive in transversely asymptotic%ally flat black string spacetimes.%%   

\section{Summary}
Finally, in combination with the previous work, we will mention the sufficient conditions for the positivity of the Y-ADM mass density to hold black strings whose boundary condition satisfie transverse asymptotically flatness.  

{\it The Y-ADM mass density becomes positive if a transversely asymptotically flat black string spacetime which contains an apparent horizon and satisfies the dominant energy condition follows the below conditions ;\\
\\
(1) either conformally flat cases or algebraically special spacetimes, or there is a translationary Killing field along the black string, i.e. the black string is uniform and
(2) the apparent horizon becomes null.}

\section*{Acknowledgements}
 
We thank Associated Professor K. Nakao and Professor H. Ishihara for useful comments. We would like to thank Associated Professor T. Shiromizu for continuous encouragement.

\end{document}